\documentclass[aps,showkeys]{revtex4}
\usepackage{epsfig,graphics}
\begin{document}
\title{Short-time dynamic exponents of an Ising model with competing
interactions}

\author{Nelson Alves Jr. and Jos\'e Roberto Drugowich de Fel\'{\i}cio},
\affiliation{Faculdade de Filosofia, Ci\^{e}ncias e Letras de
Ribeir\~{a}o Preto, Universidade de S\~{a}o Paulo, 14040--901,
Ribeir\~{a}o Preto, SP, Brazil}
\date{\today}

\begin{abstract}
In this work the two-dimensional Ising model with nearest- and
next-nearest-neighbor interactions is revisited. We obtain the dynamic
critical exponents $z$ and $\theta$ from short-time Monte Carlo
simulations. The dynamic critical exponent $z$ was obtained from the
time behavior of the ratio $F_2=\langle M^2\rangle_{m_0=0}/\langle
M\rangle^2_{m_0=1}\sim t^{d/z}$, whereas the non-universal exponent
$\theta$ was estimated from the time correlation of the order
parameter $\langle M(0)M(t)\rangle\sim t^{\theta}$, where $M(t)$ is
the order parameter at instant $t$, $d$ is the dimension of the system
and $\langle (\cdots)\rangle$ is the average of the quantity
$(\cdots)$ over different samples. We have also obtained the static
critical exponents $\beta$ and $\nu$ by investigating the time
behavior of the magnetization.
\end{abstract}
\maketitle



\section{Introduction}

The two-dimensional Ising model with nearest- and
next-nearest-neighbor interactions $J_1$ and $J_2$, respectively,
presents non-universal critical behavior for $-J_2/J_1>1/2$
\cite{barber79,landau80,oitmaa81,deoliveira86,tanaka92,minami93a}. The
existence of a non-universal critical line for $-J_2/J_1>1/2$ was
suggested for the first time by van Leeuwen in 1975
\cite{vanleeuwen75}, who investigated the fixed point structure and
critical surface of two-dimensional Ising models. Subsequently, in
1977, Krinsky and Mukamel \cite{krinsky77} conjectured that the
critical behavior of the model should belong to the universality class
of the two-component vector model. Also in 1977, Nightingale
\cite{nightingale77} worked on finite size calculations of the model
and showed that the exponents vary continuosly with the coupling ratio
between nearest- and next-nearest-neighbor interactions.

The model is defined by the Hamiltonian
\begin{equation}
\mathcal{H}=-J_1\sum_{\langle i,j\rangle}S_iS_j-J_2\sum_{\langle\langle
i,j\rangle\rangle}S_iS_j,
\label{hamiltonian}
\end{equation}
where $S_i=\pm 1$ are Ising spin variables and the notation $\langle
i,j\rangle$ and $\langle\langle i,j\rangle\rangle$ denotes that each
sum runs over nearest- and next-nearest-neighbors, respectively. At
$T=0$ the ordering is ferromagnetic for $-J_2/J_1<1/2$, whereas it is
super-antiferromagnetic (SAF) for $-J_2/J_1>1/2$ \cite{fan69}. In the
SAF phase, a row (column) of up spins alternates with a row (column)
of down spins, as depicted in Fig. \ref{fig1}. From both numerical
\cite{landau80,lima00} and analytical \cite{fan69,nauenberg74}
approaches, it was shown that in the plane of the reduced temperature
$kT/J_1$ against the competition parameter $-J_2/J_1$, the phase
diagram presents ferromagnetic, paramagnetic and
super-antiferromagnetic phases. The critical line between the
ferromagnetic and paramagnetic phases belongs to the universality
class of the two-dimensional Ising model, whereas the critical
exponents along the critical line separating the SAF ordered phase
from the the paramagnetic phase are found to be non-universal. Static
critical exponents on the SAF-Paramagnetic critical line were obtained
by Monte Carlo simulations \cite{binder80,landau80}, high temperature
series expansion \cite{oitmaa81}, renormalization group calculations
\cite{deoliveira86,li01}, finite-size scaling \cite{nightingale77},
cluster variation \cite{tanaka92} and coherent-anomaly methods
\cite{tanaka92,minami93}. It was observed that such critical exponents
vary as a function of the coupling ratio $-J_2/J_1$, thus confirming
non-universality in this model.

Today it is well known that, far from equilibrium, the short-time
relaxation of the order parameter follows a universal scale form
$M(t)\sim m_0t^{\theta}$ \cite{janssen89}, where $M(t)$ is the order
parameter at instant $t$ (measured in Monte Carlo Steps per Spin -
MCS), $m_0=M(0)$ is a (small) initial order parameter value and
$\theta$ is a dynamic critical exponent associated to the anomalous
increasing of the magnetization after the quenching of the
system. Short-time Monte Carlo simulations of the model performed in a
recent paper by Ye {\em et. al} \cite{ye00} showed that the dynamic
critical exponent $\theta$ is also non-universal along the
SAF-Paramagnetic transition line, once it depends on the ratio
$-J_2/J_1$.

To our knowledge, for this model there are no previous attempts to
obtain the dynamic critical exponent $z$, which is defined as
$\tau\sim\xi^z$, where $\tau$ and $\xi$ are time and spatial
correlation lengths, respectively. Therefore, in this work we obtain
the critical dynamic exponent $z$ of the model from the time evolution
of the ratio \cite{dasilva02}
\begin{equation} 
F_2=\frac{\langle M(t)^2\rangle_{m_0=0}}{\langle M(t)\rangle^{2}_{m_0=1}}\sim
t^{d/z},
\label{f2}
\end{equation}
where $M(t)$ is the order parameter at instant $t$, $d$ is the
dimension of the system and $\langle (\cdots)\rangle$ is the average
of the quantity $(\cdots)$ over different samples with initial order
parameter value $m_0$. This ratio has proven to be useful in
determining the exponent $z$, according to recent results obtained for
the 2D Ising model, $q=3$ and $q=4$ states Potts models
\cite{dasilva02} and at the tricritical point of the Blume-Capel model
\cite{dasilva02a}. In this technique curves of $\ln(F_2)$ against
$\ln(t)$ lay on the same straight line for different lattice sizes,
without any re-scaling in time, resulting in more precise estimates
for $z$. 

In this work we also reobtain the dynamic exponent $\theta$ along the
non-universal critical line using the time correlation of the order
parameter
\cite{tome98}
\begin{equation}
\langle M(0)M(t)\rangle\sim t^{\theta},
\label{timecorrel}
\end{equation}
and compare our results with those obtained in ref. \cite{ye00}, where
the exponent $\theta$ was obtained from the scale form $M(t)\sim
m_0t^{\theta}$. The advantage in the use of Eq. (\ref{timecorrel}) is
that one does not need to fix precisely the initial order parameter
value $m_0$. The only requirement is that $\langle
m_0\rangle=0$. Contrarily, the scale form $M(t)\sim m_0t^{\theta}$
demands sharply prepared initial states with a precise value of $m_0$,
besides a delicate limit $m_0\rightarrow 0$.

We have also calculated the static critical exponents $\beta$ and
$\nu$ of the model. Such exponents are obatined indirectly because,
from numerical simulations, we are able to obtain only the ratios of
exponents $\beta/\nu z$, $1/\nu z$ and $\lambda=(d-2\beta/\nu)/z$, as
discussed in Ref. \cite{zheng98}. In order to obtain the former ratio
$(\beta/\nu z)$, we have used the scale relation
\cite{zheng98}
\begin{equation}
M(t)\sim t^{-\beta/\nu z},
\label{bdnz}
\end{equation}
according to which the order parameter $m_0=M(0)=1$ decays in the
short-time evolution of the system. On the other hand, the ratio
$1/\nu z$ was obtained from the scale relation \cite{zheng98}
\begin{equation}
\partial_\kappa\ln M(t,\kappa)|_{\kappa =0}\sim t^{1/\nu z},
\label{1dnz}
\end{equation}
which is obtained by differentiating the quantity $\ln M(t,\kappa)$ in
relation to $\kappa$ at the critical point, where $\kappa =
(T-T_c)/T_c$ and $T_c$ is the critical temperature.

From the scaling relation \cite{zheng98}
\begin{equation}
M^2(t)\sim t^{\lambda},
\label{mquad}
\end{equation}
we have also calculated the exponent
\begin{equation}
\lambda=\left (d-\frac{2\beta}{\nu}\right )\frac{1}{z}.
\label{lambda}
\end{equation}

The exponent $\nu$ is given by
\begin{equation}
\nu=\frac{1}{(1/\nu z)(z)},
\label{ni}
\end{equation}
whereas the exponent $\beta$ is given by
\begin{equation}
\beta=\left (\frac{\beta}{\nu z}\right )\left(\frac{1}{\nu z}\right)^{-1}.
\label{beta}
\end{equation}
	
Finally, from the exponents $z$, $\nu$ and $\lambda$, the critical
static exponent $\beta$ may also be obtained from
\begin{equation}
\beta=\frac{\nu}{2}(d-z\lambda).
\label{beta1}
\end{equation}

The layout of this paper is as follows: In section \ref{stmcs} we
define the order parameter of the model and we give details about the
numerical simulations performed. In section \ref{r} we present the
results obtained for both dynamic and static critical
exponents. Finally, in section \ref{sd} we briefly discuss the main
results of this work.


\section{Short-time Monte Carlo simulations}
\label{stmcs}

Simulations were performed in two-dimensional lattices with periodic
boundary conditions. The spin states were updated using one-spin-flip
heat-bath algorithm. For each value of the competition parameter
$-J_2/J_1>1/2$ considered, we used the corresponding critical
temperature given in ref. \cite{ye00}.

\begin{figure}
\epsfig{file=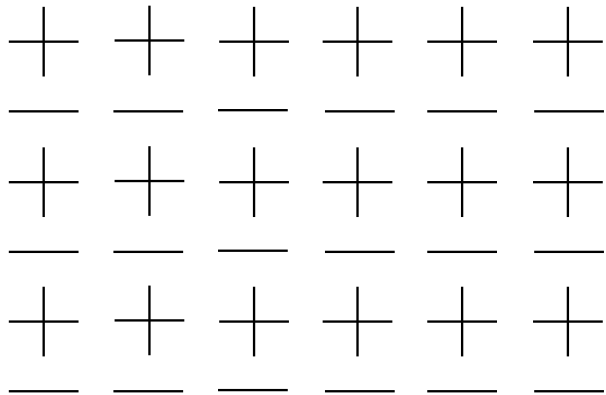,width=8cm}
\caption{Example of one possible super-antiferromagnetic (SAF)
ordering. In this example, rows of up spins alternate with rows of
down spins. The other possible configuration of this ordered state
corresponds to columns of up spins alternating with columns of down
spins.}
\label{fig1}
\end{figure}

Once the simulations were performed along the line between the SAF and
Paramagnetic phases, the order parameter $M$ considered in this work
is that corresponding to the SAF ordering. The SAF phase may be seen
as two interpenetrating sublattices, each sublattice been formed by
alternating rows, as the example depicted in Fig. \ref{fig1}: One
sublattice (say, sublattice A) corresponds to rows $+++++$, whereas
the other sublattice (say, B) corresponds to rows $-----$. By
inverting the signs of all spins in each sublattice, one obtains
another configuration in which sublattice A is formed by down spins
and sublattice B corresponds to up spins. Thus, this ground state is
twofold. We call attention of the reader to the existence of another
twofolded SAF ordering, which consists of columns of up spins
alternating with columns of down spins. Thus, the SAF ground state is
fourfold, what implies that the transition line SAF-Para is of kind
4$\rightarrow$1. Following Zittartz
\cite{zittartz81}, this phase transition is candidate to exhibit
non-universal behavior.

The order parameter $M$ is defined as $M=(M^A - M^B)/2$, where
\begin{eqnarray}
M^A&=&\frac{1}{N^2}\sum_{x=1}^{N}\sum_{y=1}^{N}S_{2x-1,y}, \\
\label{saforder-a}
M^B&=&\frac{1}{N^2}\sum_{x=1}^{N}\sum_{y=1}^{N}S_{2x,y}. 
\label{saforder-b}
\end{eqnarray}

In order to deal with square sublattices, we used lattices of $2N=144$
rows and $N=72$ columns. In the absence of a magnetic field, the
symmetry $M^A=-M^B$ is expected. Thus, when preparing initial states
for numerical simulations we have taken this expected symmetry into
account, {\em i.e.}, we prepared initial $t=0$ configurations such
that $M^A(0)=-M^B(0)$. For the exponent $z$, one sees from
Eq. (\ref{f2}) that two independent runs are necessary in order to
obtain the ratio $F_2$: One of them starts with $m_0=M(0)=1$ and the
other run starts with $m_0=M(0)=0$. For $m_0=1$, we just set
$S_{x,y}=1$ in sublattice A and $S_{x,y}=-1$ in sublattice B. On the
other hand, for $m_0=0$, we set $S_{x,y}=1$ in sublattice A and
$S_{x,y}=-1$ in sublattice B, inverting randomly the sign of half of
the spin variables in each sublattice. The ratios of exponents
$\beta/\nu z$ and $1/\nu z$ are obtained from Monte Carlo simulations
with initially ordered states, {\em i.e.}, initial states with
$m_0=1$. For the exponents $\theta$ and $\lambda$, we must have for
the initial order parameter $\langle m_0\rangle =0$. In order to
obtain such an initial state, we give a probabitlity $p=0.5$ for every
spin in sublattice A to point up. Thus, we calculate $M^A(0)$ and
impose $M^B(0)=-M^A(0)$. The spin configuration in sublattice B is
chosen in order to satisfy such imposition. Finally, the time interval
used for the runs performed in this work was [1,150], and the final
results for the dynamic critical exponents were extracted from a time
interval $[t_1,t_2] \ \ \ (1\leq t_1<t_2\leq 150)$ which maximizes the
goodness of fit $q$. For example, suppose that from two time intervals
[$t_1,t_2$] and [$t_1^{\prime},t_2^{\prime}$] one obtains $q=0.99$ and
$q=0.4$, respectively. The critical exponent obtained from the time
interval [$t_1,t_2$] is assumed to be correct.

The error bars of the exponents $\theta$, $z$, $\beta/\nu z$ and
$\lambda$, showed in the following section, are the standard deviation
from the mean value obtained from five independent runs with different
initial seeds for the random numbers generator. For the quantity
$\beta/\nu$ and for the exponents $\beta$ and $\nu$, which are
obtained indirectly, the error bars were evaluated from usual error
propagation.


\section{Results}
\label{r}

\subsection{Dynamic critical exponents}

\begin{figure}
\begin{center}
\epsfig{file=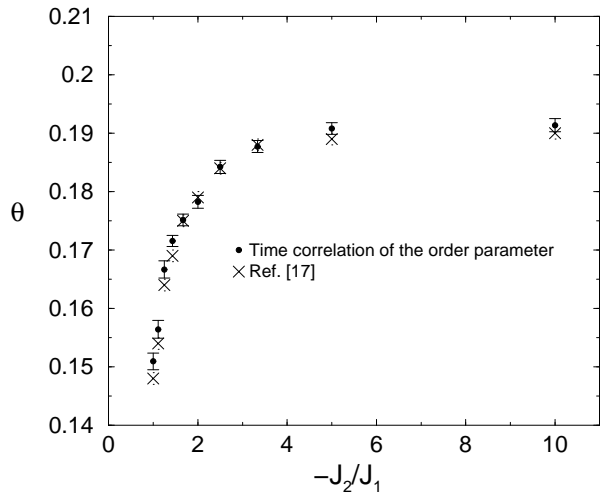,width=8cm}
\end{center}
\caption{Non-universal behavior of the dynamic critical exponent
$\theta$, as a function of the ratio between nearest- and
next-nearest-neighbor interactions $-J_2/J_1$. In the graph, the high
of each X corresponds to the error bars found in Ref. \cite{ye00}. The
points with error bars shown explicitly are the results obtained in
this work from the time correlation of the order parameter.}
\label{fig2}
\end{figure}

\begin{figure}
\begin{center}
\epsfig{file=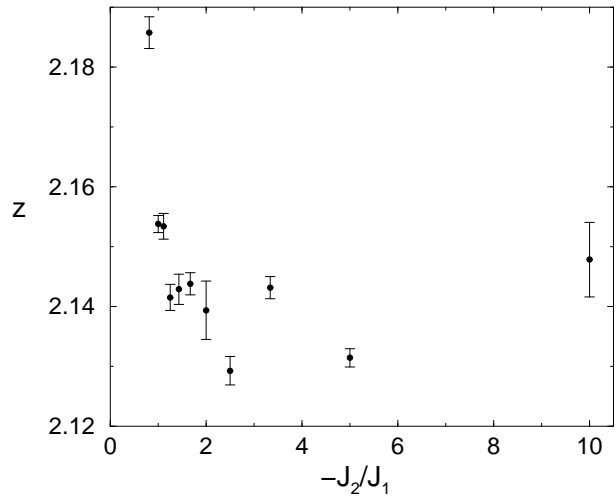,width=8cm}
\end{center}
\caption{Dynamic critical exponent $z$, as a function of the ratio
between nearest- and next-nearest-neighbor interactions $-J_2/J_1$.}
\label{fig3}
\end{figure}

In this subsection we present the results obtained for the dynamic
critical exponents $\theta$ and $z$. In Fig. \ref{fig2} it is
shown the dynamic exponent $\theta$ in function of the ratio
$-J_2/J_1$, obtained from Eq. (\ref{timecorrel}). Non-universality is
observed, since $\theta$ clearly depends on the ratio $-J_2/J_1$. This
result is in accordance with that obtained in Ref. \cite{ye00}.
In Fig. \ref{fig3} we present the dynamic exponent $z$ as a function
of $-J_2/J_1$, obtained from the ratio given in Eq. (\ref{f2}). From
Fig \ref{fig3} one can not affirm if the exponent $z$ is universal or
not, once the variation of $z$ with the coupling parameter $-J_2/J_1$
is not so pronounced as observed for the exponent
$\theta$. Furthermore, since the major variation (or fluctuation) of
$z$ occurs close to the disorder point ($-J_2/J_1=0.5$), such
fluctuation may result from a crossover effect.

From both Fig. \ref{fig2} and Fig. \ref{fig3} we observe that, as
the coupling parameter $-J_2/J_1$ increases, the exponents $\theta$ and
$z$ tend towards their known Ising values $0.19\cdots$ and
$2.15\cdots$, respectively. 
\clearpage

\subsection{Static critical exponents}

\begin{figure}
\begin{center}
\epsfig{file=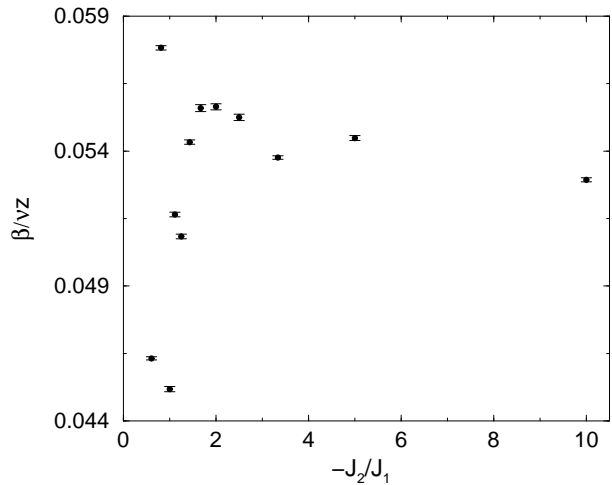,width=8cm}
\end{center}
\caption{Ratio $\beta/\nu z$, as a function of the ratio between
nearest- and next-nearest-neighbor interactions $-J_2/J_1$.}
\label{fig4}
\end{figure}

\begin{figure}
\begin{center}
\epsfig{file=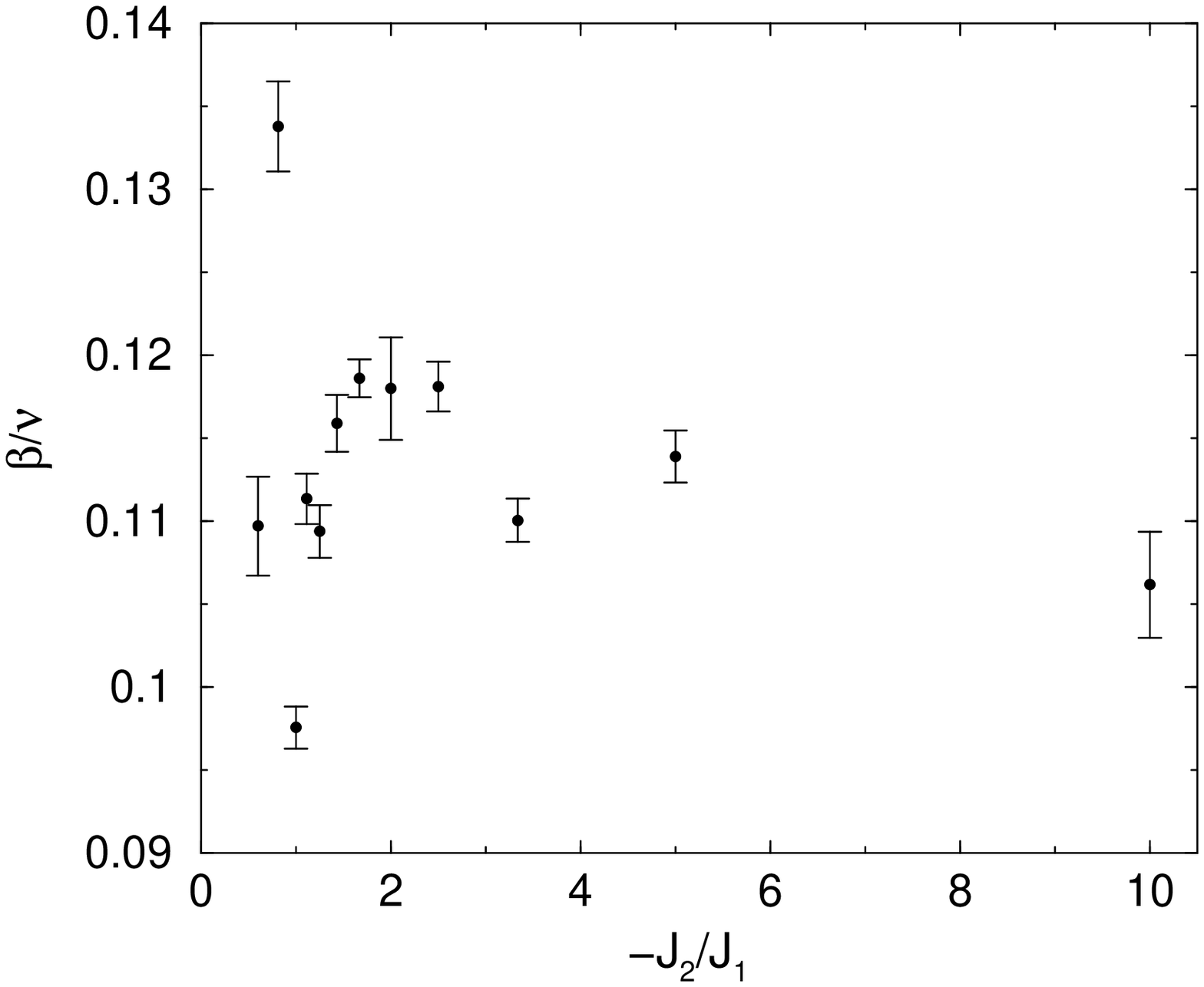,width=8cm}
\end{center}
\caption{Ratio $\beta/\nu$, as a function of the ratio between nearest-
and next-nearest-neighbor interactions $-J_2/J_1$.}
\label{fig5}
\end{figure}

\begin{figure}
\begin{center}
\epsfig{file=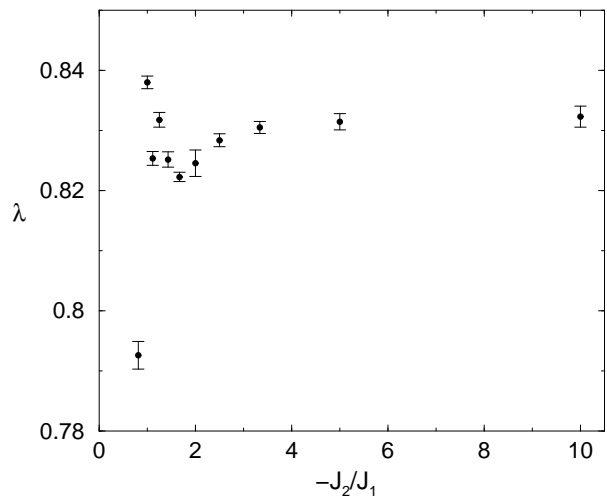,width=8cm}
\end{center}
\caption{Exponent $\lambda=(d-2\beta/\nu)/z$, as a function of the
ratio between nearest- and next-nearest-neighbor interactions
$-J_2/J_1$.}
\label{fig6}
\end{figure}

In this subsection we show, as a function of the ratio $-J_2/J_1$, the
ratio of the exponents $\beta/\nu z$, $\beta/\nu$ and
$\lambda=(d-2\beta/\nu)/z$, and the static critical exponents $\beta$
and $\nu$. In Figs. \ref{fig4}, \ref{fig5} and \ref{fig6},
as a function of the coupling parameter, we show the ratios $\beta/\nu
z$, $\beta/\nu$ and $\lambda$. We observe that the behavior of these
quantities as the coupling parameter $-J_2/J_1$ varies is quite
similar to that observed for the exponent $z$. So, we are not able to
affirm anything about non-universality of these
quantities. Particularly, from Fig. \ref{fig5} we observe that the
ratio $\beta/\nu$ does not reach its known exact Ising value
$\beta/\nu =1/8$ for large enough values of $-J_2/J_1$. We confirmed
this result by Finite Size Scaling calculations combined with
conformal invariance \cite{alves02}.

\begin{figure}
\begin{center}
\epsfig{file=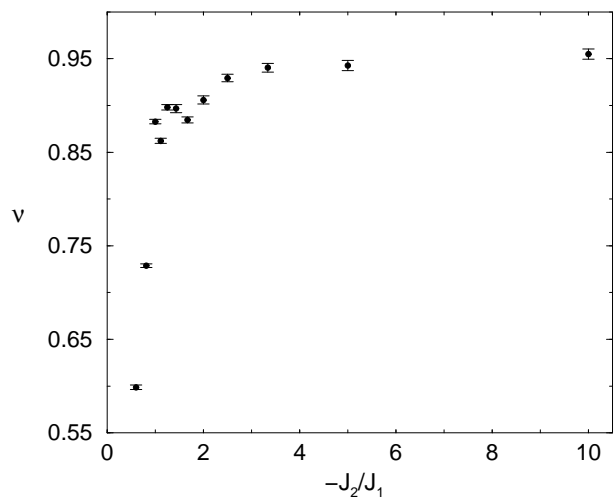,width=8cm}
\end{center}
\caption{Non-universal behavior of the static critical exponent $\nu$,
as a function of the ratio between nearest- and next-nearest-neighbor
interactions $-J_2/J_1$.}
\label{fig7}
\end{figure}

The static critical exponent $\nu$, obtained from Eq. (\ref{ni}), is
shown in Fig. \ref{fig7} as a function of the coupling parameter
$-J_2/J_1$. The non-universal behavior observed for this exponent was
already found before using Monte Carlo \cite{landau85} and
transfer-matrix \cite{nightingale77} methods. In Fig. \ref{fig7} we
observe that the exponent $\nu$ tends towards the known exact
two-dimensional Ising value $\nu =1$ as $-J_2/J_1$ increases.

\begin{figure}
\begin{center}
\epsfig{file=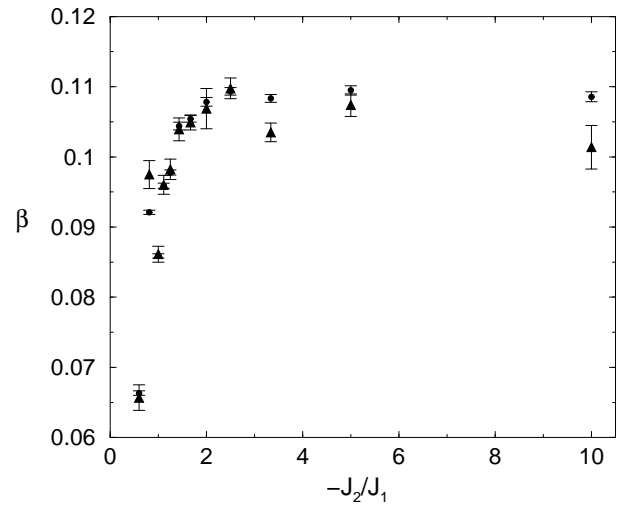,width=8cm}
\end{center}
\caption{Non-universal behavior of the static critical exponent
$\beta$, obtained from Eqns. (\ref{beta}) (full circles) and
(\ref{beta1}) (triangles up), as a function of the ratio between
nearest- and next-nearest-neighbor interactions $-J_2/J_1$.}
\label{fig8}
\end{figure}

The non-universal behavior of the static critical exponent $\beta$,
evaluated from both Eqs. (\ref{beta}) and (\ref{beta1}), is shown in
Fig. \ref{fig8} as a function of the ratio $-J_2/J_1$. We observe
the good agreement for the numerical values obtained for the exponent
$\beta$ from Eqns. (\ref{beta}) and (\ref{beta1}). In the second case,
we note bigger error bars. This is due to the fact that the evaluation
of the exponent $\beta$ from Eq. (\ref{beta1}) involves more exponents
previusly obtained, and so the error propagation implies the bigger
error bars observed. 
\clearpage

\section{Summary and Discussion}
\label{sd}

In this work we study the dynamic critical exponents of the
two-dimensional Ising model with competing nearest- and
next-nearest-neighbors interactions. We obtained the dynamic exponent
$z$ along the transition line between the SAF and Paramagnetic phases
by using a method recently proposed based on the mixing of initial
conditions \cite{dasilva02}. To our knowledge, this work is the first
attempt to obtain the exponent $z$ for this model. We have also
calculated the dynamic critical exponent $\theta$ using the time
correlation scale form of the order parameter \cite{tome98} and our
results are in good agreement with those obtained in ref. \cite{ye00}.
Non-universality was also observed for the static critical exponents
$\beta$ and $\nu$. In particular, as the ratio $-J_2/J_1$ increases,
we observed that the dynamic critical exponents tend towards their
known Ising values. However, the ratio $\beta/\nu$ does not tends
towards its known exact Ising value $\beta/\nu =1/8$ as the coupling
ratio $-J_2/J_1$ increases. We have confirmed this result by means of
Finite Size Scaling caculations combined with conformal invariance
\cite{alves02}.

\begin{acknowledgments}
The authors acknowledge financial support from Brazilian agencies
FAPESP and CAPES. The authors are also grateful to Departamento de
F\'{\i}sica e Matem\'atica (DFM) at IFUSP for computer facilities.
\end{acknowledgments}
\newpage
\bibliography{abbrev,innni2d}
\bibliographystyle{apsrev}

\end{document}